# THERMOCURRENTS AND THEIR ROLE IN HIGH Q CAVITY PERFORMANCE

R. Eichhorn[#], C. Daly, F. Furuta, A. Ganshyn, D. Gonnella, D. Hall, V. Ho, G.H. Hoffstaetter, M. Liepe, J. May-Mann, T. O'Connell, S. Posen, P. Quigley, J. Sears, and V. Veshcherevich
CLASSE, Cornell University, Ithaca, NY 14853, USA

*Abstract*

Over the past years it became evident that the quality factor of a superconducting cavity is not only determined by its surface preparation procedure, but is also influenced by the way the cavity is cooled down. Moreover, different data sets exists, some of which indicate that a slow cool-down through the critical temperature is favourable while other data states the exact opposite. Even though there were speculations and some models about the role of thermo-currents and flux-pinning, the difference in behaviour remained a mystery.

In this paper we will for the first time present a consistent theoretical model which we confirmed by data that describes the role of thermo-currents, driven by temperature gradients and material transitions. We will clearly show how they impact the quality factor of a cavity, discuss our findings, relate it to findings at other labs and develop mitigation strategies which especially address the issue of achieving high quality factors of so-called nitrogen doped cavities in horizontal test.

## INTRODUCTION

Continuous wave mode operation of future accelerators like ERLs or the LCLS-II [1] haven driven the research on achieving high quality factor SRF cavities to keep operation cost low. As the surface resistance of superconducting cavities approach the theoretical limits parasitic effects limiting the performance came into focus of research. One interesting finding was that the quality factor of a cavity is impacted by the cool-down rate.

This effect was first reported by HZB [2] where the authors concluded that a slow cool-down results in a smaller surface resistance and higher Q of the cavity. Similar results were gained at Cornell, seeing that an initial cool-down to 4 K, followed by a thermo-cycle warming to 20 K and a slow re-cool through the critical temperature increased the quality factor significantly[3]. In contrast to this, FNAL [4] saw an increase of the quality factor of nitrogen doped cavities after a fast cool-down. These contradictory findings have been reproduced and confirmed by others so there is no doubt about the validity of the data.

Thermo-currents have been a candidate to explain these findings for several years. Even though their existence was not in question, their impact on the cavity performance was never clearly understood. Recent experiments at Cornell from vertical and horizontal tests have now revealed a breaking in symmetry, allowing a consistent explanation of the data, including the deterioration of the quality factor.

## BACKGROUND

During cavity testing inside the Cornell HTC in the framework of our ERL R&D we confirmed a very interesting effect: The quality factor Q of a superconducting cavity can be increased after the initial cool-down by going through a second cool-down cycle which warms up the cavity to 15 -20 K and then slowly cools it back to 2 K again, transiting the critical temperature with a gradient as low as 0.5 K/h.

By this cycle, we were able to increase the Q at 1.8 K from $3.5 \cdot 10^{10}$ to $6 \cdot 10^{10}$. This has already been reported in [5]. In the absence of having a good understanding of this effect, several possible explanations were investigated.

To check if the shielding efficiency of the magnetic shielding enclosing the cavity (and ensuring an appropriate damping of the earth magnetic field in the vicinity of the cavity) is deteriorated during the cool-down, we started to equip our test with a cryogenic flux gate magnetometer that measured the magnetic field close to the cavity inside the inner magnetic shielding [6]. Interestingly enough, we found that the magnetic field changes drastically during cool-down: it started from an ambient value of 0.15 µT rising to 0.45 µT, then going down and reversing its direction (which needed a manual adjustment, resulting in some missing data, see fig. 9 at 19:00 h) to -0.25 µT to become -0.11 µT as the cavity transits through the critical temperature becoming superconducting. This behaviour, depicted in fig. 1, is clearly not compatible with a temperature dependent efficiency of the magnetic shielding.

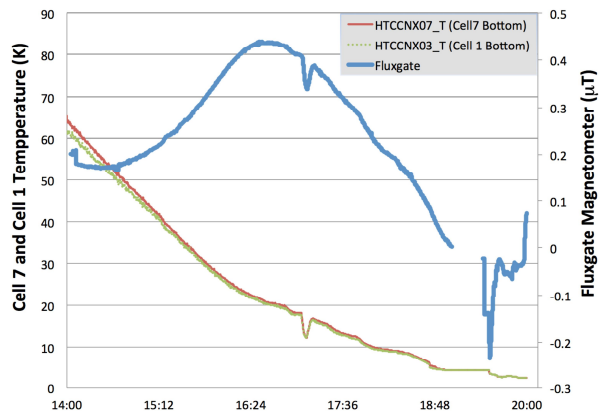

Figure 1: Magnetic field measurement during initial cool-down of a cavity in a horizontal test. The flux gate magnetometer was mounted outside the helium tank but inside the second layer of magnetic shielding, measuring only one (arbitrary) component of the magnetic field.

___________________
[#]r.eichhorn@cornell.edu

One potential explanation of this finding is magnetic fields, induced by thermo-currents. Today's SRF cavities are made from niobium which becomes superconducting at 9.2 K. Cooled by liquid helium, these cavities are usually enclosed by a vessel made from titanium, welded to the cavity at the cut-off tubes. With both material transitions held at different temperatures, there is the potential to drive a persistent thermo-current.

However, arguments were made that thermo-currents should not lead to any magnetic field outside the helium vessel (where the magnetometer in our set-up was located) nor inside the cavity (where it could affect the quality factor by means of flux trapping) [7,8].

The arguments were based on symmetry, solving Maxwell's equations analytically. We recently found conditions, under which this symmetry is broken that completely changes the perspective under which thermo-currents have to be seen.

When analysing data from different sources one should be aware that cavity testing can also be conducted in a bath cryostat where the cavity is not necessarily enclosed by a Ti-vessel. Even though vertical tests are usually conducted with bare cavities while horizontal test require cavities with Ti-vessel, the effect from horizontal or vertical cooling has to be clearly distinguished, as we describe below.

## SEEBECK EFFECT

Thermo-currents are the result of the Seebeck-Effect, which is well known in physics for more than a century: Discovered in 1826, Seebeck found that a current will flow in a closed circuit made of two dissimilar metals when the two junctions are maintained at different temperatures. However, the effect also exists within a single, uniform metal, where a voltage potential builds up between the warmer and the colder end of the material.

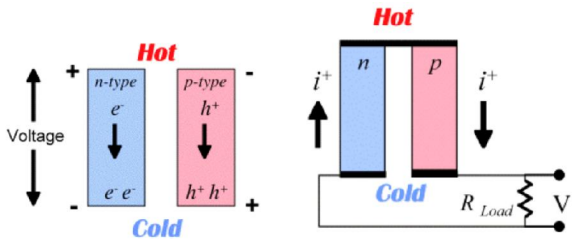

Figure 2: Visualisation of the Seebeck voltage of different metals (taken from [9]), not connected (left) and connected on one side (right). In the non-connected case a static voltage builds up resulting in a vanishing current in the equilibrium. If a loop exist (connected scenario), a persistent current is excited even in the equilibrium state.

Table 1: Thermoelectric power (seebeck coefficient) for niobium and titanium, taken from [10]. Values are given in µV/K.

|    | 10 K | 20 K | 50 K  | 80 K  | 100 K |
|----|------|------|-------|-------|-------|
| Nb | 0.31 | 0.98 | 2.73  | 3.09  | 3.13  |
| Ti | N/D  | N/D  | -3.00 | -3.00 | -2.60 |

The polarity and the value of that emf voltage is dependant on the material, leading to the definition of the Seebeck coefficient S:

$$\Delta U = S \cdot \Delta T. \quad (1)$$

Seebeck coefficients of metals can have either sign as they are defined relative to platinum. In a single metal arrangement, depicted by fig. 2 (left), this voltage exists across the metal but does not result in a current flowing other than simply building up the charges, initially. If there is a material transition, where two different metals are joined, not only does a potential difference exist, but it might also drive a persistent current (driven by the temperature difference) if the loop is closed (see fig.2, right diagram).

As superconducting cavities are made out of niobium while the helium vessel enclosing them is typically titanium this effect is relevant for accelerator physics: During the cool-down of a dressed cavity (a cavity welded into its helium vessel) it is easy to imagine that both ends of the cavity (where the Nb-Ti transition is located) have different temperatures.

The emf voltage following the Seebeck theory is then given by

$$U_{th} = (S_{Nb} - S_{Ti}) \cdot \Delta T. \quad (2)$$

To complicate the story, Seebeck coefficients are also temperature dependent. The data for niobium and titanium, taken from [10] are given in table 1. Below 50 K the thermoelectric power of titanium is unknown. As the Seebeck effect vanishes for all materials at zero temperature we assume in our analysis a linear dependency of the coefficient with temperature between 0 K and the first data point at 50 K. With the more general definition of the Seebeck coefficient, the thermo-voltage becomes

$$U_{th} = \int_{T_1}^{T_2} (S_{Nb}(T) - S_{Ti}(T)) dT \quad (3)$$

## THERMO-CURRENT TEST SET-UP

To investigate the Seebeck effect in detail and especially to confirm that the thermoelectric power of titanium is different from that of niobium below 50 K, we built a window-frame set-up, as shown in fig. 3. It simulates a cavity/ he-vessel arrangement by having two transitions between niobium and titanium. Each transition was equipped with a cernox thermometer. While the lower end was immersed in liquid helium and the upper end was heated. The Seebeck voltage, which is expected (eqn. (3) and table 1) to be in the order of µV, drives a current, which due to the low resistance of the circuit, given by

$$R = \frac{\rho L}{A} \quad (4)$$

is potentially in the order of 10-100 Ampere.

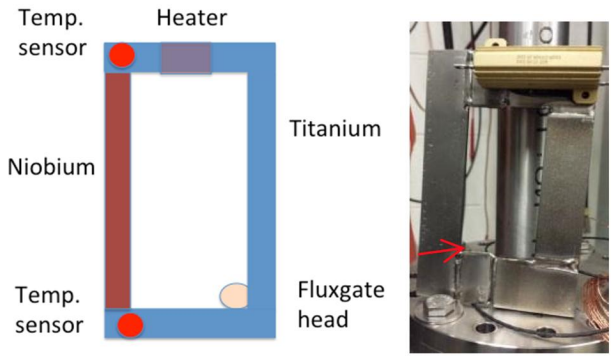

Figure 3: Set-up to measure the thermo-currents. On the left is a sketch of the arrangement with the principle components, on the right is a photograph of the set-up.

Even with this rather high current there was no direct way to measure it without requiring additional material transitions. We decided to measure the current indirectly by its generated magnetic flux. A fluxgate sensor was placed in the corner of the window-frame and the analysis was guided by Biot-Savart's law

$$d\vec{B} = \frac{\mu_0}{4\pi} \frac{Id\vec{l} \times \vec{r}}{r^2} \quad (4)$$

The result of the measurement is given in fig. 4. As the heater in the set-up influenced the magnetic reading due to its (small but visible) stray field we only considered data taken without current through the heater to be relevant. By that, we found an almost linear dependency from the temperature difference- as theory predicts for a constant Seebeck coefficient. This linear dependency will become relevant later when we discuss cavity performance.

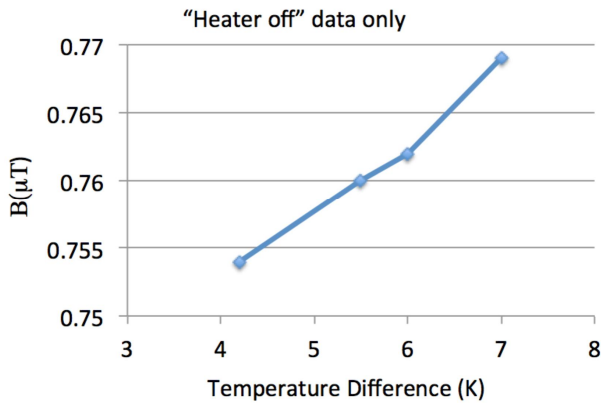

Figure 4: Measured magnetic field as induced by the thermo-current. The extrapolated field for ΔT approaching 0 K is the residual magnetic field inside the cryostat. For details see text.

## VERTICAL CAVITY TEST DATA

In the framework of our ERL program [11] we built 6 cavities for the Main Linac Cryomodule (MLC) prototype [12]. All of them were tested vertically before the helium vessel was welded to the cavity. In order to understand the thermo-current effect, we tested the cavity fully insulated from its support frame. A second measurement was done after the helium vessel was welded to the cavity which then allows thermo-currents to flow. The data that has been taken is shown in fig. 5 and 6.

We conducted several test going through the critical temperature slowly and fast, the results of which are shown in fig. 5. The graph gives the magnetic field at the iris of one cavity cell, measuring the azimuthal magnetic field, which is the expected orientation of a thermo-current induced field. Both datasets where taken on dressed cavities.

Beside the stronger fields observed on the fast cool-down there is an important difference in the data: the field remaining in the iris area was 0.1 µT for the slow cool-down but as high as 0.53 µT after a fast cool-down. Concluding that this is the frozen flux seems to be too bold, as the kink in the data close to transition temperature reveals a rather complicated dynamics which we are currently investigating in more detail. However, it can be stated that higher fields are present on cool-down for a fast cycle, pointing again to thermo-currents given the fact that on the fast cool-down the temperaturespread between the bottom and the top of the cavity is greater.

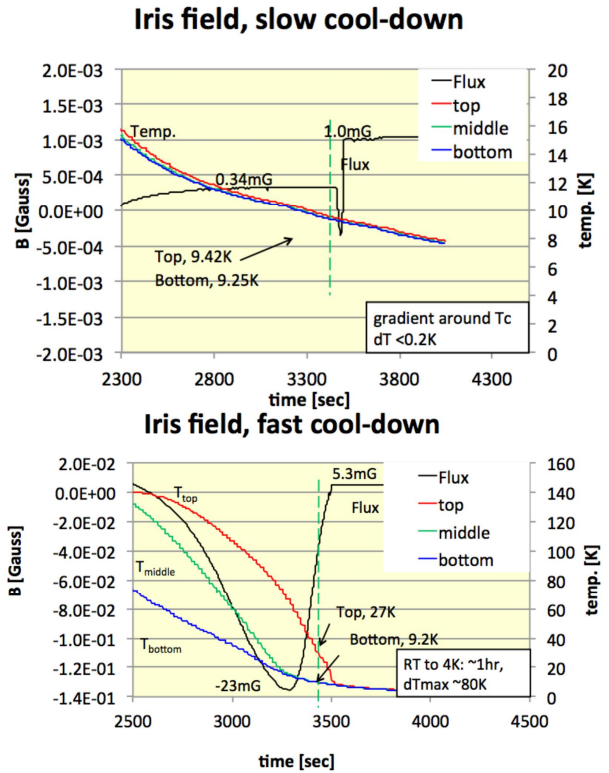

Figure 5: Slow (upper plot) and fast (lower plot) cool-down cycle: shown are temperatures of the cavity and the magnetic flux measured at the iris location.

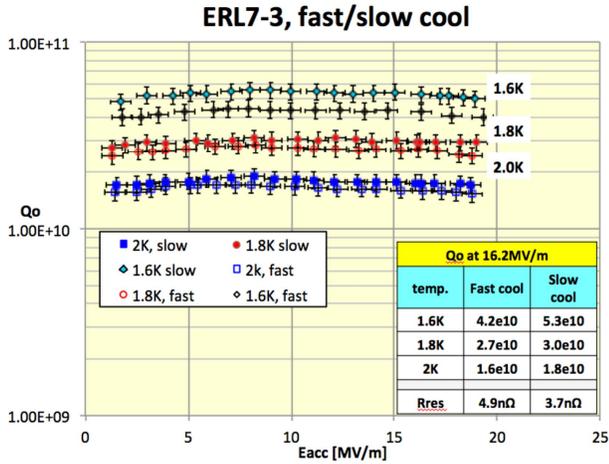

Figure 6: Cavity Qs of the ERL7-3 cavity (dressed) for different temperatures and different cool-down cycles. Consistently, a reduction of 1 nΩ in the residual surface resistance has been found for a slow cool-down.

After each cool-down cycle, we measured the quality factor of the cavities. Figure 6 summarizes the results of one of the cavities (ERL7-3). It should be noted that the cavity was not removed from the dewar between the tests. Depending on the cool-down cycle we found different Q factors with Qs higher for a slow cool-down. As the effect increases with decreasing temperature we conclude that a slow cool-down impacts the residual resistance, only.

Even though symmetry-based arguments suggest that the cavity inner surface is not affected by the magnetic field of a thermo-current [7] and thus the quality factor/ residual resistance is unaffected, too, we confirmed by several measurements a decrease of 1 nΩ in the residual resistance after a slow cool-down. It should be noted that these results confirm results from a conventionally treated cavity [13] and contradicts findings on nitrogen-doped cavities [14,15].

As pointed out in [8] a small deviation in the concentric alignment of the cavity/ vessel assembly can lead to a symmetry breaking which is able to explain these results. However, we will show that there is a different mechanism leading to a symmetry breaking which is able to even explain a stronger impact of the cool-down cycle on the residual resistance, which we observed in later tests.

## HORIZONTAL TEST DATA

Cryomodule test within the LCLS-II high $Q_0$ program [14] allowed us to study cool-down effects on a nitrogen-doped cavity. For this test, a nitrogen-doped cavity was welded into an ILC style titanium helium vessel, and installed in the Cornell Horizontal Test Cryomodule (HTC)[15]. Magnetic field probes and temperature sensors on the cells of the cavity allowed for detailed measurements of temperature gradients and magnetic fields during cavity cool-down, including a direct measurement of the thermoelectric induced magnetic field. Figure 7 shows the location of these sensors. A heater was placed on one of the beam tubes of the cavity to generate large longitudinal gradients during cool down, potentially resulting in a large thermo-voltage, high thermo-currents and increased induced magnetic field during transition to superconductivity.

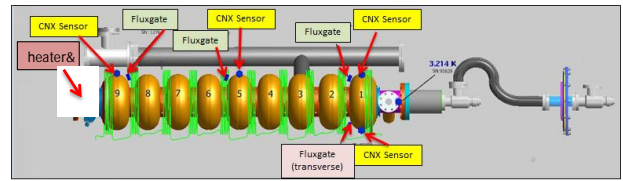

Figure 7: Horizontal test set-up to measure a nitrogen doped cavity, indicating the location of the instrumentation and diagnostics.

Figure 8 shows the magnetic field, measured by the magnetometer placed between the cavity and the helium vessel. Beside the small residual magnetic field, one clearly sees the onset of the thermo-electric induced field, becoming noticeable at a temperature difference of about 10 K between the cavity ends. It should be noted that every data point requires a warm-up and cool-down which explains the limited data set. However, this is again, clear evidence proving the existence of thermo-currents.

To understand the impact on the cavity performance, we measured the quality factor of the cavity after each cool-down. The data is reported in fig. 9 for the situation were no or a negligible thermo-current is excited (ΔT= 6.8 K) as well as for a medium and high thermal gradient situation (ΔT= 21.8 K and 29.4 K). The quality factors we got where $2.7 \cdot 10^{10}$ ($\Delta T$= 6.8 K), $2.0 \cdot 10^{10}$ ($\Delta T$= 21.8 K) and $1.5 \cdot 10^{10}$ ($\Delta T$= 29.4 K). Values quoted refer to 2 K and 16 MV/m.

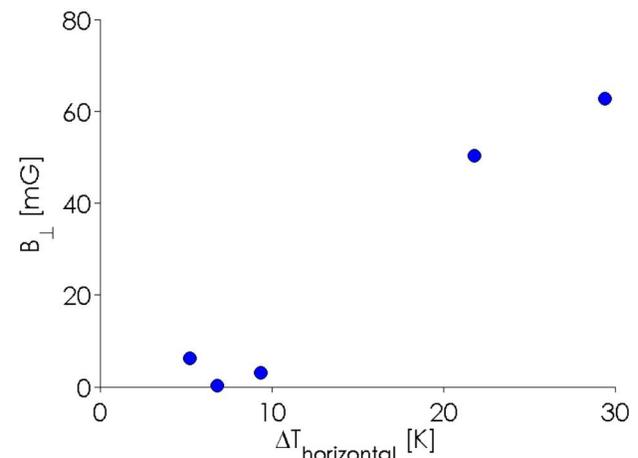

Figure 8: Transverse magnetic field measured by the fluxgate magnetometer, located between the helium vessel and the cavity outer wall as a function of the temperature difference between the cavity's ends. Data was taken with the Cornell HTC near $T_c$.

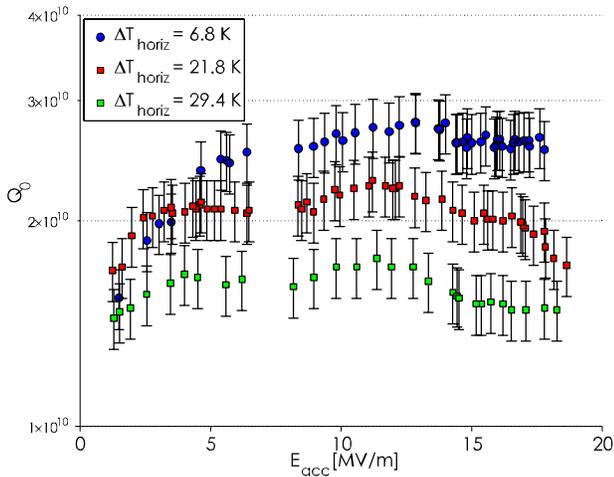

Figure 9: $Q_0$ vs $E_{acc}$ performance of an LCLS-II N-doped 9-cell cavity at 2.0 K in the Cornell Horizontal-Test Cryomodule for different temperature gradients between the cavity ends near $T_c$. Uncertainty on $E_{acc}$ is 10%.

All measurements were done subsequently on the same cavity so a change in the intrinsic quality factor (which we see as the BCS and the none-cool down-related residual portion) can be excluded.

Decomposing the quality factor into its components leads to the plot in fig. 10 where the residual resistance of the cavity surface is plotted against the magnetic flux measured during transition. It should be noted that the data seems to be no longer linear correlated which might be due to differences in flux pinning and cool-down speed variations.

So far, we still have to show why the thermo-current induced magnetic fields that we measured between the cavity outer wall and the helium vessel impacts the RF surface of the cavity. This will be explained in the next section.

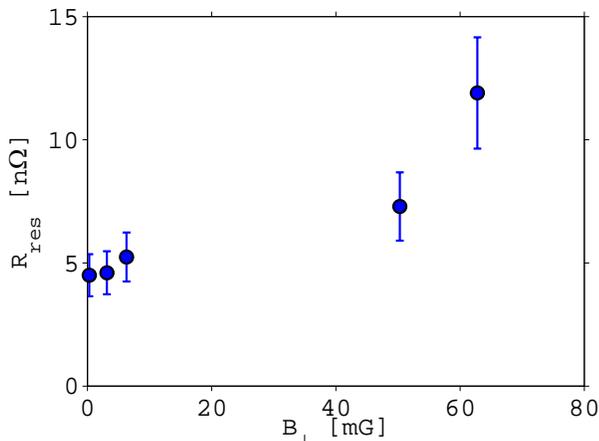

Figure 10: Residual surface resistance as a function of the measured transverse magnetic field (induced by thermoelectric currents in the Cornell HTC).

## THERMOCURRENT SIMULATION

Previously, analytical analysis argued that the axial symmetry of SRF cavities leads to no (or when considering the potential asymmetry from vessel or cavity port negligible small) thermo-electric induced magnetic fields in the relevant RF penetration layer at the inner cavity surface [8]. They concluded that therefore thermo-electric currents are not a concern for the performance of SRF cavities. However, our findings indicated early-on that thermo-electric currents may have a more severe impact on the SRF performance as so far predicted.

In order to gain a better understanding, numerical simulations with CST® EM-Studio® were undertaken. We modelled a real size cavity with a simplified helium vessel (see fig. 11). The Seebeck voltage was applied over an artificial gap on the right side of the helium vessel, depicted by the ports visible in fig 13 and 14. For the simulation, realistic values for the expected thermo-voltage and the resistivity of the materials were used [10,17] and the mesh was carefully adjusted to avoid numerical problems.

In order to understand the results, a distinction into two different cases is appropriate:

### Azimuthally symmetric case

This scenario is depicted in fig. 11: even though the cavity/ helium-vessel may have non-uniform properties, symmetry exists if the properties are independent of the azimuth. In this scenario, a thermo-current is excited if a temperature gradient exists along the z-axis: The disparity of the temperatures at the material transitions results in a Seebeck voltage, driving this current.

However, due to the postulated symmetry, currents in the upper and the lower half are equal, resulting in a magnetic field that only exist between the outer cavity wall and the helium vessel. As a consequence of the vanishing magnetic field at the RF surface of the cavity, thermo-currents in this symmetric case do not result any in contribution to the flux pinning at transition.

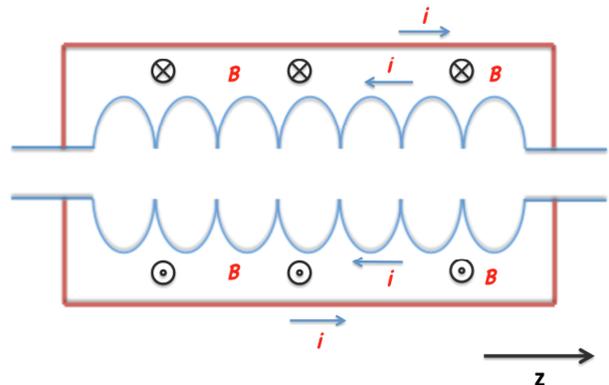

Figure 11: Thermo-currents and associated magnetic field in the azimuthal symmetric scenario: the Seebeck induced currents i are equal, producing a symmetric magnetic field.

*Non-Azimuthal symmetric case*

As described, the temperature gradient along the z-axis generates the Seebeck voltage. It relates to the current over the resistivity of the circuit. If no azimuthal symmetry exists, the resistivity for example in the upper half might be higher compared to the lower half. In that case the Seebeck voltage would result in an unevencurrent distribution. As a result, the magnetic field distribution would also be asymmetric and magnetic fields inside the cavity will exist, as shown in fig. 12. It is easy to imagine that during the cool-down, not only a temperature gradient between the cavity ends exists (resulting in the Seebeck voltage) but also a radial or transversal gradient (top to bottom) which diverts the currents asymmetrically as a result of temperature dependent resistance. In the extreme case one might assume that part of the cavity is already superconducting while the remaining portion is still resistive.

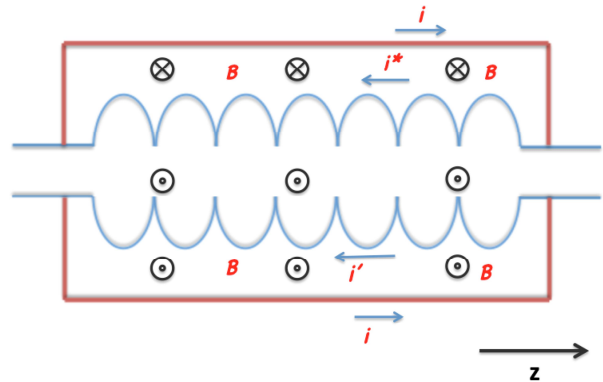

Figure 12: Thermo-currents and associated magnetic field in the asymmetric scenario. In this case, the currents along the helium vessel are approximately equal (explained in the text) but i* and i' might not be equal, resulting in a field configuration with fields inside the cavity.

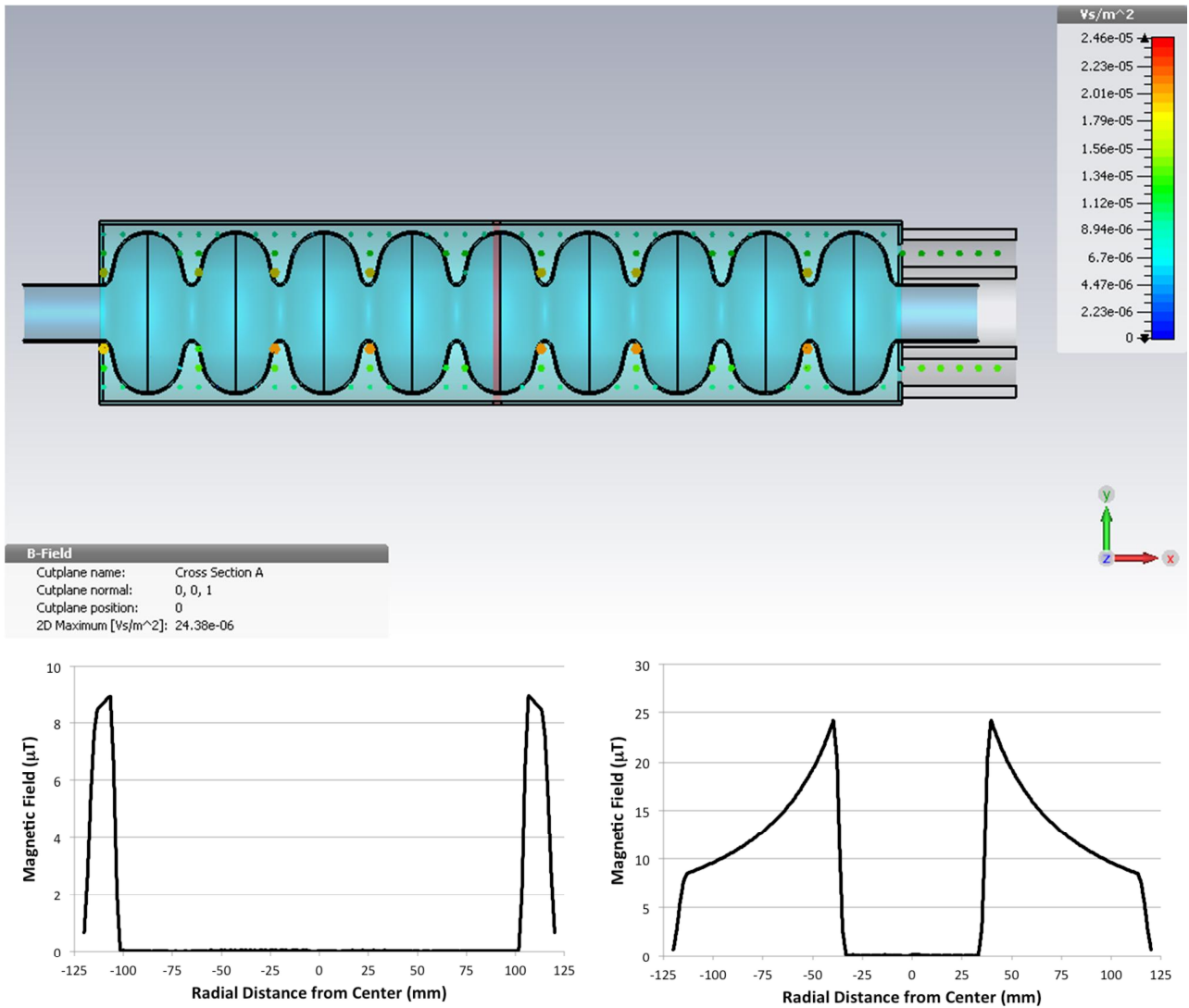

Figure 13: Results of the numerical field simulation with the parameters given in the text, assuming azimuthal symmetry. The top graph shows the 3-D field configuration, plots below give z-axis cuts along one equator (bottom left) and at an iris (bottom right), both locations where close to the cavity centre. As expected, the magnetic field resulting from the thermo-current fully vanish inside the cavity.

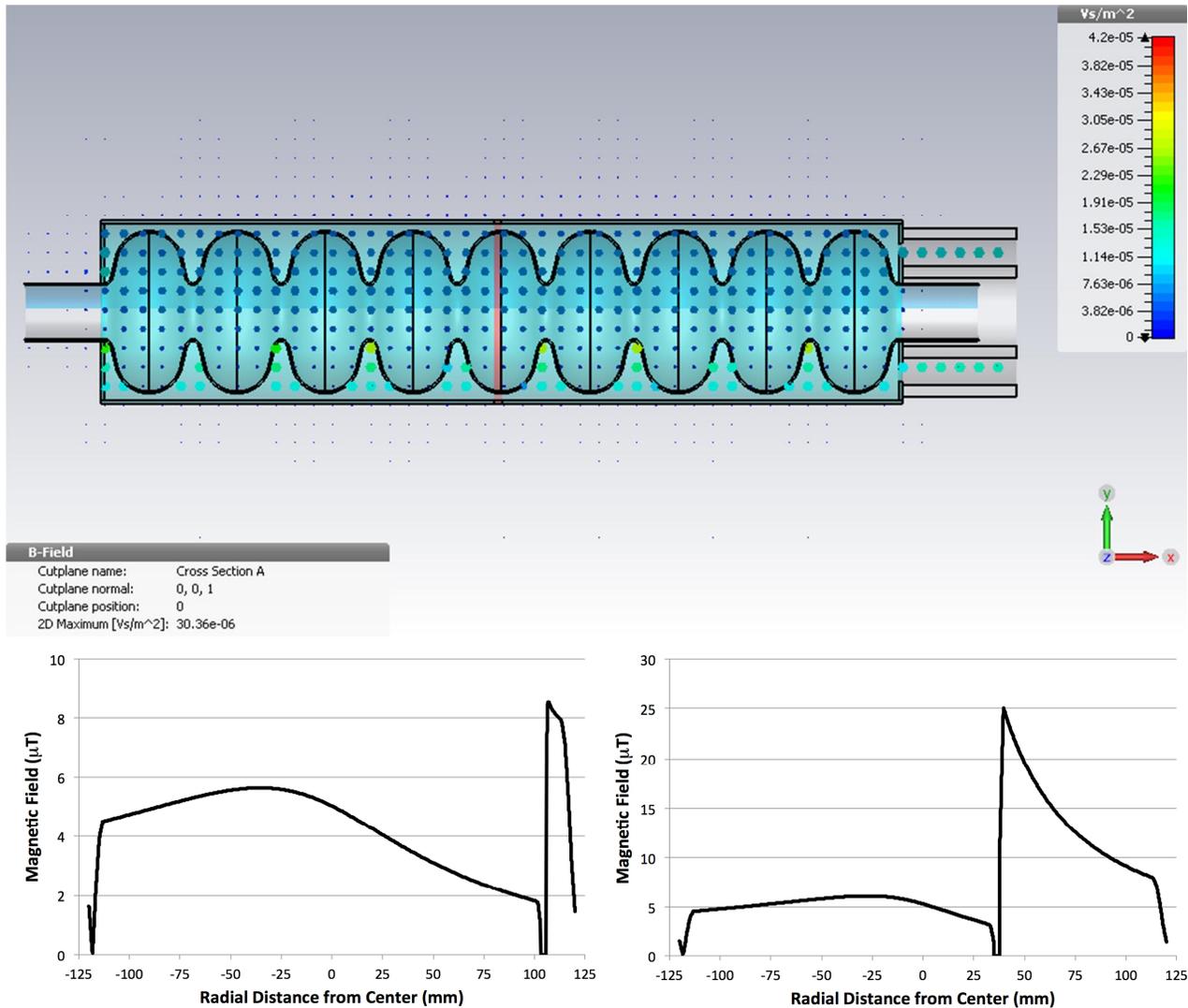

Figure 14: Results of the asymmetric calculation, where the lower portion of the cavity is assumed to be superconducting while the upper half remains normal conducting. Again, the upper graph shows the 3-D field configuration which now is asymmetric, too. The z-axis cuts (bottom, equator cut on the left, iris data on the right) reveal a significant amount of magnetic flux inside the cavity which at transition to superconductivity of the upper half is susceptible to pinning and as a result increasing the surface resistance.

*Results*

For the numerical simulation we assumed a Seebeck voltage of 150 µV, which corresponds to a temperature of 10 K on one side and 50-60 K on the other side (the calculation is based on experimental conditions as published in [14,15] and the linear extrapolation described above). We also assumed constant (which can be interpreted as a mean) resistivity for the niobium ($5 \cdot 10^{-10}$ Ωm) and the titanium ($2.5 \cdot 10^{-7}$ Ωm). The bellow of the titanium vessel was accounted for this simulation in terms of resistance but it was not modelled geometrically. We calculated the current in the thermo- loop to be 4.8 A and the maximum magnetic field to be 25 µT. The results for the symmetric scenario are shown in fig. 13, where the upper plot gives the magnetic field configuration. The plots below give the magnetic field along a z-axis cut at the location of the equator (left) and the iris (right). As expected, fields are symmetric and no field inside the cavity exist.

It should be noted that this plot also explains the different magnetometer readings we got related to the thermo-currents, depending on their positioning at the iris (where the field is enhanced) and the equator, where only a weaker field is measurable.

To simulate the asymmetric scenario we assumed the same conditions as above with only one difference: we characterized the lower portion of the cavity to be a perfect conductor- representing its vanishing resistance in the superconducting state. The field configuration yielded is given in fig. 14 (upper plot), the lower plots are z-cuts at the iris and equator, respectively. As a result of the azimuthal asymmetry, magnetic fields are asymmetric and a reasonable large magnetic field exists inside the cavity which during transition of the upper half of the cavity through $T_c$ could be trapped, causing an increase in the residual resistance and thus deterioration in the cavity Q. It is remarkable that the field strength at the RF surface, being 5-10 µT is in good agreement with the measured field (fig. 5).

Figure 14 also indicates that a measurable thermo-current induced magnetic field is generated outside the titanium vessel. This explains our initial findings referred in fig. 1. Given the field configuration has been simulated, this allows to index the field at the cavity surface without having to place a magnetometer inside the cavity. Thus, a magnetic field reading outside the helium vessel permit a direct distinction between the themo-electric magnetic fields which do not affect performance (symmetric, no field inside the cavity nor outside the helium vessel) an the fields which impact the performance (asymmetric with field inside the cavity and outside the helium vessel).

*Interpretation*

If an azimuthal asymmetry exists, thermo-currents can generate magnetic field at the RF layer of the cavity that is subject to flux trapping. A reason for this asymmetry can be found in the cool-down, if a transversal temperature gradient in the dressed cavity exists. This is usually the case in a horizontal test, where the cavity is cooled through a stream of cold helium entering through a cool-down port at the lower portion of the helium vessel, while the exhaust is located on the top. This results in a lower temperature of the lower portion of the cavity with decreased resistance and increased current in that region. As the resistivity of the titanium is almost constant below 50 K [17], the change of resistance has to be caused by the niobium with the most drastic change to happen as the niobium becomes superconducting.

The thermo-current effect has less influence in vertical tests for two reasons. Usually, only bare cavities are tested, but a closed current loop might exist over the cavity support frame. Longitudinal temperature gradients might be huge resulting in large thermo-current induced magnetic fields. However, due to the mostly preserved azimuthal symmetry as a result of the only z-dependant temperature distribution, fields are symmetric eventually generating no flux at the RF layer of the cavity.

## TRANSITION DYNAMICS

The transition dynamics of a dressed cavity in a horizontal test seems to be a complicated process. As the cavity cools down, the Seebeck voltage increases due to the thermoelectric power of the two materials involved. At the same time the loop resistance decreases. Both effects lead to an increase in the current and induced magnetic field, as show in fig. 5, lower plot, up to t = 3300 s (The decrease in field after t = 3300 s is due to the decrease in the temperature difference).

However, as the niobium becomes superconducting several more effects take place: first, the resistance of the loop will drop slightly, increasing the current. In the asymmetry scenario we calculated an increase of 5-10%. However, the Seebeck voltage would eventually drop rapidly as a superconductor by definition does not have thermoelectric power - so the net effect might be a decrease in current and fields.

On top of that, the niobium begins to expel or pin flux. Characterized by the change in the permeability and the magnetization that goes with it, the field configuration changes. Overall, this makes it difficult to judge, which of these effects contribute to which amount to the spike, seen in fig. 6, upper plot at t = 3400 s (which also exists in the lower plot at t = 3500 s). More detailed investigations are currently being conducted.

## MITIGATION STRATEGIES

Our analysis shows that in order to minimize thermo-currents, any temperature gradients should be avoided. However, there is evidence that the cool-down has also an effect on the amount of magnetic flux trapped as the cavity becomes superconducting [18-20]. It seems, that a fast cool-down leads to less pinning. The increase of surface resistance caused by flux pining on N-doped cavities is more severe compared to conventional cavities [20]. As it stands, high cool-down rates might be required if the shielding of the cavity against residual magnetic field is insufficient. The trade-offs between the optimum shielding, cool-down rate and ratio of symmetry have to be assessed by additional studies.

It is yet not fully clear if the higher flux expulsion on fast cool-downs is related to a thermal gradient or due to the temperature change rate. This will be subject to further investigations.

From what we know, one can state that a longitudinal temperature gradient drives the thermo-current induced field but the transversal gradient determines how much of this flux hits the sensitive RF surface of the cavity. However, there seems to be no way to avoid a transversal gradient in a horizontal set-up which means one should avoid longitudinal gradients. This is optimally achieved by having two cool-down lines feeding the both ends at the bottom of the helium vessel and a centrally located gas exhaust. Both cool-down lines have to have a well balanced helium mass flow to achieve a minimal temperature difference of the two material transitions on either end.

But there exists two more mitigation options. Our analysis showed that the resistance of the cavity slightly above Tc is ~2.5 µΩ, while the helium vessel has ~30 µΩ. Half of that resistance is coming from the bellow in the helium vessel, necessary to allow tuning of the cavity. Doubling the bellow's length would increase the resistance of the thermo-current loop by 50% and as the Seebeck-voltage is determined by the thermal gradient, the resulting current and magnetic fields would decrease to 67 %. Together with a well-balanced cooling scheme minimizing longitudinal gradients this might resolve the issue of trapping thermo-current induced magnetic fields.

A rather rigorous mitigation strategy would be replacing the titanium helium vessel around the cavity by a niobium container. In lacking a material transition, thermo-voltages in this arrangement would never lead to a loop current. It should be mentioned that some of the low beta cavities operated successfully (see for example [21]) have helium vessels made from reactor grade niobium with a negligible impact on overall costs.

## CONCLUSIONS

We have investigated the effect of thermo-currents in dressed cavities and their impact on the quality factor. We demonstrate the existence of magnetic fields associated with the currents and proved their contribution to the performance, which is minor in vertical testing but can be severe in horizontal test. Our model allows a systematic explanation of the findings which points to a longitudinal temperature difference to drive the current while the transversal gradient determines the amount of asymmetry which results in generating fields at the surface layer. The asymmetry can be diagnosed by a magnetometer placed outside the helium vessel. Based on this, trade-off studies on magnetic shielding and cool-down procedures can be conducted.

## ACKNOWLEDGEMENT

The authors would like to acknowledge the great boost the LCLS-II high Q R&D program gives to the field which allowed us to continue investigating the mystery of thermo-currents. Special thanks go to the Cornell cryomodule team for assembling the HTC in record time and to the cryogenics group for all their support. We also like to express our gratitude to the entire FNAL SRF group for supplying a nitrogen doped 9-cell cavity.